\documentclass[12pt]{article}
\usepackage[dvips]{graphicx}
\usepackage{latexsym}
\parskip=\medskipamount

\newcommand {\cF}{{\cal F}}
\newcommand {\cG}{{\cal G}}

\newcommand {\cN}{{\cal N}}

\def\a{\alpha}
\def \bi{\bibitem}

\def\b{\beta}

\def\G{\Gamma}

\def\j{\psi}

\def\l{\lambda}
\def\m{\mu}

\def\o{\omega}
\def\p{\pi}
\def\q{\theta}

\def\t{\tau}

\def\x{\xi}
\def\z{\zeta}
\def\D{\Delta}
\def\F{\Phi}
\def\J{\Psi}
\def\L{\Lambda}
\def\O{\Omega}

\def\U{\Upsilon}

\newcommand{\ad}{{\dot{\alpha}}}                           %new
\newcommand{\bd}{{\dot{\beta}}}                            %new
\newcommand{\hf}{\frac12}
\newcommand{\vf}{\varphi}
\newcommand{\be}{\begin{equation}}
\newcommand{\ee}{\end{equation}}
\newcommand{\bea}{\begin{eqnarray}}
\newcommand{\eea}{\end{eqnarray}}
\newcommand{\non}{\nonumber}
\begin{document}
%%%%%%%%%%%%%%%%
%%%%%%%%%%%%%%%%

\begin{titlepage}
\thispagestyle{empty}

\begin{flushright}
hep-th/0403082 \\
March,  2004 \\
\end{flushright}
\vspace{5mm}

\begin{center}
{\Large \bf
Relaxed super self-duality and 
effective action
}
\end{center}
%\vspace{3mm}

\begin{center}
{\large S. M. Kuzenko and   I. N. McArthur}\\
\vspace{2mm}

\footnotesize{
{\it School of Physics, The University of Western Australia\\
Crawley, W.A. 6009, Australia}
} \\
{\tt  kuzenko@cyllene.uwa.edu.au},~
{\tt mcarthur@physics.uwa.edu.au} \\
\end{center}
\vspace{5mm}

\begin{abstract}
\baselineskip=14pt
A closed-form expression is obtained for 
a holomorphic sector of the two-loop 
Euler-Heisenberg type effective action 
for $\cN=2$ supersymmetric QED derived
in hep-th/0308136.  In the framework of 
the background-field method, 
this sector is singled out by computing
the effective action for a background 
 $\cN=2$ vector multiplet satisfying a relaxed 
super self-duality condition.
The approach advocated in this letter
can be applied, in particular, 
to the study of the $\cN=4$ super Yang-Mills theory 
on its Coulomb branch.
\end{abstract}

\vfill
\end{titlepage}

\newpage
\setcounter{page}{1}

\renewcommand{\thefootnote}{\arabic{footnote}}
\setcounter{footnote}{0}
%%%%%%%%%%%%%%%%%%%%%%%%%%%
%\sect{Introduction}

The two-loop extension of the (one-loop)
 QED Euler-Heisenberg 
action \cite{EH,W,Schwinger} was derived by 
Ritus \cite{Ritus} almost thirty years ago.
Unlike the original (one-loop) Euler-Heisenberg action, 
its two-loop  extension \cite{Ritus} 
involves a rather complicated 
double proper-time integral.
It has recently been  demonstrated 
\cite{DSch}  that for a self-dual background
the two-loop QED effective action  \cite{Ritus} 
takes a remarkably simple form, 
and actually becomes very similar 
to the one-loop action 
 \cite{EH,W,Schwinger} in the same background.

The Euler-Heisenberg Lagrangian
corresponds to an approximation 
of  slowly varying fields, and  
is a function of the field strength $F_{ab}$ only, 
\be
L_{\rm EH} =L(F^2_+ , F^2_- )~.
\ee
Here $F_+$ and $F_-$ are 
the (anti) self-dual components  
of the field strength $F$, 
\bea 
F_\pm = \hf (F &\mp &{\rm i}\, \tilde{F})~,
\qquad \widetilde{F_\pm } = \pm \,{\rm i} \, F_\pm ~, \non \\
F_\pm^2 = -\hf (\cF & \mp &{\rm i} \, \cG )\,{\bf 1}_4~,
\qquad \cF= {1\over 4} F^{ab}F_{ab}~, \quad 
\cG= {1\over 4} F^{ab}\tilde{F}_{ab}~,
\eea
with $\tilde{F}$ the Hodge-dual of $F$.
The general structure of $L_{\rm EH} $ is 
as follows: 
\be
L (F^2_+ , F^2_- ) = 
\L(F^2_+  ) +
{\bar \L} ( F^2_- )
~+~  F^2_+  F^2_- \,
\O ( F^2_+ , F^2_- )~. 
\label{EH}
\ee
 Choosing a self-dual background, 
$F_- =0$, one can keep track of the 
holomorphic function $\L$, 
but all the information about  
the function $\O$ is lost.

In supersymmetric  theories,
there are no quantum corrections 
to $\L$ and $\bar \L$ 
beyond second order
in the field strength, $\L \propto F_+^2$, 
since no appropriate superfield invariant exists.
In other words,   
the effective Lagrangian is essentially  
trivial in the case of self-dual fields
(see \cite{Grisaru,CDDG,DI,Kallosh} 
for an incomlete list of references), 
and therefore the
results of \cite{DSch} cannot be 
applied  directly.  Fortunately, we can 
still profit, although
rather indirectly, from the approach 
advocated in \cite{DSch}. 
The point is that
the function $\O$ in (\ref{EH})
has the following general form 
\be
\O ( F^2_+ , F^2_- ) = \o ( F^2_+ )
+{\bar \o} ( F^2_- ) 
~+~  F^2_+  F^2_- \,
\U ( F^2_+ , F^2_- )~.
\ee
Its holomorphic part, 
$ \o ( F^2_+ ) $, 
can be restored by computing 
the effective action for background
vector supermultiplets satisfying 
a relaxed self-duality condition.
 
In superfield notation, 
a relaxed super self-duality condition 
can be defined by 
\be
W_\a \neq 0~, \quad 
D_\a W_\b = 0~, \qquad 
{\bar D}_{(\ad} {\bar W}_{\bd)} \neq 0
\label{N=1-relaxed}
\ee
in the case of $\cN=1$ supersymmetry, or
\be 
D^i_\a W  \neq 0~, \quad 
D_\a^i  D_\b^j W = 0~, \qquad 
{\bar D}^i{}_{(\ad} {\bar D}_{\bd) i}  {\bar W} \neq 0
\label{N=2-relaxed}
\ee
in  the case of $\cN=2$ supersymmetry.
Here $W_\a$ and $W$ are the chiral 
superfield strengths describing the $\cN=1$ 
and $\cN=2$ Abelian vector multiplets, 
respectively.
Ordinary (Euclidean) super self-duality \cite{Zumino}
corresponds to  setting 
$W=0$ while keeping 
$\bar W$ non-vanishing
(see also \cite{Kallosh,Siegel}).  From the point 
of  view of $\cN=1$ supersymmetry, the $\cN=2$ 
vector multiplet strength 
$W$ consists of two $\cN=1$ superfields:
(i) a chiral scalar $\F$; and (ii) the $\cN=1$ 
vector multiplet strength $W_\a$.
The conditions on $W_\a$ which follow from 
 (\ref{N=2-relaxed}) coincide with 
(\ref{N=1-relaxed}). 

The condition of relaxed super self-duality 
has a simple meaning at the component level. 
In the case of an off-shell $\cN=2$ vector multiplet,
its chiral strength $W$ is known to contain 
the following component fields
(with $U| \equiv U(x,\q)|_{\q=0}$):
(i) a complex scalar $\vf = W |$;
(ii) two left-handed spinors 
$\j^i_\a =D^i_\a W |$; 
(iii) a symmetric bi-spinor
$F_{\a \b} = D^i_{(\a } D_{\b ) i} W |$
which is in one-to-one correspondence with $F_-$;
(iv) an auxiliary iso-triplet 
$X^{ij} = D^{\a (i } D^{j)}_\a W|$.
The relaxed super self-duality 
requires $F_{\a \b} = X^{ij} =0$ 
and allows for  non-vanishing 
$\vf$ and $\j^i_\a$. 
This is clearly a relaxation of the ordinary 
super self-duality requirements
$\vf = \j^i_\a =F_{\a \b} = X^{ij} =0$.

In Minkowski space-time, 
the conditions (\ref{N=1-relaxed})
and (\ref{N=2-relaxed}) are purely formal, as they 
are obviously inconsistent 
with the structure of a single real vector multiplet.
Nevertheless, their use is completely legitimate
if we are  only interested  in computing some
special, holomorphic-like  sector of the effective
action. To be more specific, let us consider $\cN=2$ 
supersymmetric  QED (SQED).

The action of $\cN=2$ 
SQED written in terms of $\cN=1$ 
superfields is 
\bea
S_{\rm SQED} &=& \frac{1}{e^2} \int {\rm d}^8 z \, {\bar \F} \F
+  \frac{1}{e^2} \int {\rm d}^6 z \, W^\a W_\a \non \\
&& +\int {\rm d}^8 z \, \Big( \overline{Q} {\rm e}^V Q 
+ \overline{ \tilde{Q} } {\rm e}^{-V} \tilde{Q} \Big)
+ \Big(
{\rm i}  \int {\rm d}^6 z \, \tilde{Q} \F Q + {\rm c.c.} \Big)~,
\label{n=2sqed-action0}
\eea
where $W_\a = - {1\over 8} {\bar D}^2 D_\a V$. 
The dynamical variables $\F$ and $V$ describe an
$\cN=2$ Abelian vector multiplet, while 
the superfields $Q$ and  $\tilde{Q}$ constitute  
a massless Fayet-Sohnius hypermultiplet.
The case of a massive hypermultiplet is obtained 
from (\ref {n=2sqed-action0}) by the shift
$\F \to \F+ m$, with $m$ a complex parameter.\footnote{The 
action of $\cN=1$ SQED is obtained from (\ref {n=2sqed-action0})
by discarding $\F$ as a dynamical variable, and instead
`freezing' $\F$ to a constant value $m$.}

We are interested in a low-energy effective action
 $\G[W,  \F]$ which describes the dynamics of the $\cN=2$
massless vector multiplet and which is generated  by integrating 
out  the massive charged hypermultiplet. 
More precisely, we concentrate on a slowly 
varying part of $\G[W,  \F]$  that, at the component level, 
comprises contributions with (the supersymmetrization of) all 
possible powers of the gauge field strength without derivatives.
Its generic form is \cite{BKT}
\be
\G[W,  \F] = \Big(
\a \int{\rm d}^6 z \,W^2\, \ln {\F \over \m} 
~+ ~ {\rm c.c.} \Big) 
~+~ \int{\rm d}^8 z \,{ {\bar W}^2 W^2 \over  {\bar \F}^2\F^2 } 
\, \O(\J^2, {\bar \J}^2)~,
\label{structure}
\ee
where 
\be
{\bar \J}^2 = {1 \over4} D^2 \Big( { W^2  \over  {\bar \F}^2\F^2 } \Big)~,
\qquad 
\J^2 = {1 \over4} {\bar D}^2 \Big( { {\bar W}^2  \over  {\bar \F}^2\F^2 } 
\Big)~,
\label{psi}
\ee
$\m$ is the renormalization scale and 
$\O$ some real analytic function.
The first term on the right hand side of (\ref{structure})
is known to be one-loop exact in perturbation theory
(see, e.g.,  \cite{BKO}), 
while the second term receives quantum corrections 
at all  loops \cite{BKT,KM2}.

Up to a scale transformation, the function 
$\O$ in (\ref{structure})
is the same as the one in (\ref{EH}).
To compute its holomorphic part, 
$\O(\J^2, 0)$, within the background field 
formulation, it is sufficient to evaluate 
covariant supergraphs  for the case when
$\F $ and $\bar \F$ are constant, while 
$W_\a$ and ${\bar W}_\ad$ obey 
the conditions (\ref{N=1-relaxed}).\footnote{In 
the case of $\cN=4 $ $SU(N)$
super Yang-Mills theory on its Coulomb 
branch, the choice of such a background
tremendously simplifies 
the evaluation of  the two-loop
effective action \cite{KM4}.} 
Indeed, the use of such a background 
allows one to keep track, in loop calculations, 
of the following  sector 
$$
\int{\rm d}^8 z \,{ {\bar W}^2 W^2 \over  {\bar \F}^2\F^2 } 
\, \O(\J^2, 0)
$$
of the effective action. 
Once  this functional form has been computed, 
one can remove the condition of relaxed 
super self-duality and 
work with arbitrary off-shell superfields.
At the component level, the bosonic part 
of the functional   is then 
$$
\int{\rm d}^4 x \,{ 
( F^2_+  F^2_- 
%\over  
/ \D^2) 
%({\bar \vf} \vf)^2 
} \, 
\, \O( { F^2_+ / \D^2
%\over ({\bar \vf} \vf)^2 
}
, 0)~, \qquad \D = {\bar \vf} \, \vf ~,
$$
modulo terms involving the auxiliary fields
and derivatives of $\vf$ and $ \bar \vf$.
For these reasons, relaxed super self-duality 
proves to be useful in Minkowski space.

In this paper, our attention will be restricted
to the consideration of the real part of
the effective action.  
At  the one-loop level, the function $\O$ in (\ref{structure})
is \cite{BKT,KM2} (see also \cite{MG,PB})
\bea
\O_{\rm one-loop}(\J^2, {\bar \J}^2)
= 
\frac{1 }{(4\p)^2}
\int\limits_0^\infty   {\rm d}  s \, s\,
\z(s \J,s{\bar \J}) \, 
{\rm e}^{ - s }~,
\label{one-loop}
\eea
where
\be
\z(x,y)=\z(y,x) = 
\frac{ y^2(\cosh x - 1) - x^2(\cosh y -1)} 
{x^2 y^2(\cosh x - \cosh y)}~. 
\ee
${}$From this,
\bea 
\O_{\rm one-loop}(\J^2, 0)
=  \frac{1 }{4(4\p)^2}
\int\limits_0^\infty   {\rm d}  s \, s\,
\left\{ {1 \over (s \J/2)^2 } - 
{ 1 \over \sinh^2 (s\J /2) } \right\} 
\,  {\rm e}^{ - s }~.
\label{one-loop-2}
\eea
In terms of the function \cite{DSch}
\be
\x(x) =-x \left( { {\rm d}\over {\rm d} x} \, \ln \G (x)
-\ln x +{1 \over 2x} \right)  
= \hf \int\limits_0^\infty   {\rm d}  s \,
\left\{ 
{1 \over s^2 } - 
{ 1 \over \sinh^2 s } \right\} 
\,  {\rm e}^{ - 2xs }~, 
\label{xi}
\ee
the $\O_{\rm one-loop}(\J^2, 0)$ 
is seen to be proportional to the first derivative
of $\x$. What happens at two loops?

To answer this, we turn to the two-loop effective 
action for $\cN=2$ SQED which was 
computed in \cite{KM2} 
on the base of the covariant multi-loop technique
of \cite{KM}: 
\bea
\G_{\rm two-loop} &=&
\int{\rm d}^8 z \,{ {\bar W}^2 W^2 \over  {\bar \F}^2\F^2 } 
\, \O_{\rm two-loop} (\J^2, {\bar \J}^2)~, \non \\
\O_{\rm two-loop} 
(\J^2, {\bar \J}^2)
&=& 
\O^{\rm I+II}
(\J^2, {\bar \J}^2)
+ \O^{\rm III}
(\J^2, {\bar \J}^2)~.
\eea
Here $\O^{\rm I+II}$ and $\O^{\rm III}$ 
correspond to different two-loop supergraphs
\cite{KM2}, and have the form
\bea
 \O^{\rm I+II} (\J^2, {\bar \J}^2) 
&=& \frac{e^2}{(4\p)^4 }
\int \limits_{0}^{\infty} {\rm d}s
\int \limits_{0}^{\infty}  {\rm d}t \,
{\rm e}^{-(s+t) }  
\frac{ s\l_+}{\sinh (s\l_+)} \, \frac{ s\l_-}{\sinh (s\l_-)} \,
\frac{ t\l_+}{\sinh (t\l_+)} \, \frac{ t\l_-}{\sinh (t\l_-)} \non \\
& \times &
\frac{\sinh (s\J/2)}{s\J/2} \, 
\frac{\sinh (s{\bar \J} /2)}{s {\bar \J}/2} \,
\frac{\sinh (t\J/2)}{t\J/2} \, \frac{\sinh (t{\bar \J} /2)}{t {\bar \J}/2}  
\, I_2(s,t)
~, 
\eea
and
\bea
\O^{\rm III} (\J^2, {\bar \J}^2) 
&=& \frac{e^2}{2(4\p)^4 }
\int \limits_{0}^{\infty} {\rm d}s
\int \limits_{0}^{\infty}  {\rm d}t \,
{\rm e}^{-(s+t) } 
\frac{ s\l_+}{\sinh (s\l_+)} \, \frac{ s\l_-}{\sinh (s\l_-)} \,
\frac{ t\l_+}{\sinh (t\l_+)} \, \frac{ t\l_-}{\sinh (t\l_-)} \non \\
& \times &
\left\{ \frac{\sinh^2 (s\J/2)}{(s\J/2)^2} \,  
\frac{\sinh^2 (t{\bar \J} /2)}{(t{\bar \J}/2)^2}  ~+~ 
(s \leftrightarrow t) \right\} \, I_1(s,t)~, 
\eea
where 
\be
\l_\pm =\hf ({\bar \J} \pm \J )~,
\ee
and  $I_{1}(s,t)$ and $I_{2}(s,t)$
denote the following simple proper-time
integrals
\bea 
I_1(s,t)&=& \int\limits_{0}^{\infty} \frac{{\rm d}u } {u^2} \,
\frac{1} {(u^{-1} +a_+)(u^{-1} +a_-)}  ~, 
\label{I-1} \\
I_2(s,t)&=& \int\limits_{0}^{\infty} \frac{{\rm d}u } {u^2} \,
\frac{1} {(u^{-1} +a_+)(u^{-1} +a_-)} 
\left( \frac{P_+}{u^{-1} +a_+}
+ \frac{P_-}{u^{-1} +a_-} \right) ~,
\label{I-2}
\eea
with
\be 
a_\pm = \l_\pm \coth(s\l_\pm) + \l_\pm \coth(t \l_\pm)~,
\qquad
P_\pm = 
\frac{ \l_\pm} 
{\sinh (s\l_\pm)}\,\frac{\l_\pm}{ \sinh(t \l_\pm)}~.
\ee

We are going to evaluate 
$\O_{\rm two-loop} (\J^2, 0)$.  Since
\be
{\bar \J}=0 \quad \longrightarrow \quad 
\l_\pm = \pm \hf \J  ~,
\ee
 then
\be 
a_\pm = {\J \over 2}   \,\frac{ \sinh \Big(  (s+t)\J /2\Big)}
{\sinh (s \J / 2) \sinh (t \J / 2) } ~,
\qquad 
P_\pm = 
\frac{ \J/2} 
{\sinh (s\J /2)}\,\frac{\J/2}{ \sinh(t \J/2)}~,
\ee
and therefore 
\bea 
I_1(s,t)&=& {2 \over \J} \,
\frac{ \sinh (s \J / 2) \sinh (t \J / 2) }
{ \sinh \Big(  (s+t)\J /2\Big)} ~, \\
I_2(s,t)&=& 
\frac{ \sinh (s \J / 2) \sinh (t \J / 2) }
{ \sinh^2 \Big(  (s+t)\J /2\Big)} ~.
\eea

${}$For $\O^{\rm I+II}
(\J^2, 0)$ we therefore get
\bea
 \O^{\rm I+II} (\J^2, 0) 
&=& \frac{e^2}{(4\p)^4 }
\int \limits_{0}^{\infty} {\rm d}s
\int \limits_{0}^{\infty}  {\rm d}t \,s\,t\,
{\rm e}^{-(s+t) }  \,
\frac{  (\J / 2)^2 }
{ \sinh^2 \Big(  (s+t)\J /2\Big)} ~.
\eea
The double proper-time integral here
can be reduced to a single integral, 
by introducing new integration variables,
$\a$ and  $\t $, defined as follows
(see, e.g.,   \cite{DR})
\bea 
s+t = \t~,  \quad  s-t =\t \,\a~, \quad
\qquad \t \in [0, \infty ) ~, 
\quad \a \in [-1, 1]~, 
\label{change-of-var}
\eea
such that 
\be
\int \limits_{0}^{\infty} {\rm d}s
\int \limits_{0}^{\infty}  {\rm d}t \, 
L\Big(s, t \Big ) = 
\hf \int \limits_{0}^{\infty} {\rm d}\t 
\int \limits_{-1}^{+1} {\rm d}\a \, \t \, 
L\Big( s(\a, \t), t(\a, \t) \Big) ~.
\ee
This leads to 
\bea
 \O^{\rm I+II} (\J^2, 0) 
&=& \frac{e^2}{6(4\p)^4 }
\int \limits_{0}^{\infty} {\rm d}s \, s^3\,
\frac{ (\J/2)^2 }
{ \sinh^2  (s\J /2 )} \, {\rm e}^{-s }  \non \\
&=& \frac{e^2}{6(4\p)^4 }
~+~
 \frac{e^2}{6(4\p)^4 }
\int \limits_{0}^{\infty} {\rm d}s \, s^3\,
\left\{ \frac{ (\J/2)^2 }
{ \sinh^2  (s\J /2 )} 
- {1 \over s^2} \right\}\, {\rm e}^{-s } ~.
\label{omega-I+II}
\eea

${}$For $\O^{\rm III}
(\J^2, 0)$ we obtain
\bea
 \O^{\rm III} (\J^2, 0) 
&=& \frac{e^2}{(4\p)^4 }
\int \limits_{0}^{\infty} {\rm d}s
\int \limits_{0}^{\infty}  {\rm d}t \,s^2\,
\frac{ (\J/2) \,\sinh  (t\J /2 ) } 
{ \sinh  (s\J /2 ) \, \sinh \Big(  (s+t)\J /2\Big)} \,
{\rm e}^{-(s+t) } 
 ~.
\eea
Using the identity
\be
\frac{ \sinh t } {\sinh s \, \sinh (s+t) }
= \coth s - \coth (s+t)~, 
\ee
we can rewrite $\O^{\rm III}(\J^2, 0)$ 
as follows: 
\bea
 \O^{\rm III} (\J^2, 0) 
&=& \frac{e^2}{(4\p)^4 }
\int \limits_{0}^{\infty}  {\rm d}t {\rm e}^{-t } 
\int \limits_{0}^{\infty} {\rm d}s \, s^2\,
(\J / 2) 
\, \coth (s\J/2) \, {\rm e}^{-s } \non \\
&-& \frac{e^2}{(4\p)^4 }
\int \limits_{0}^{\infty} {\rm d}s
\int \limits_{0}^{\infty}  {\rm d}t \,s^2\,
(\J / 2) \,  \coth \Big(  (s+t) \J /2\Big) \,
{\rm e}^{-(s+t) }~. 
\eea
In the first term here, one of the proper-time integrals
is elementary.  In the second term, 
the double proper-time integral can be reduced to a single
one by implementing the change of variables
(\ref{change-of-var}). 
This gives
\bea
 \O^{\rm III} (\J^2, 0) 
&=& \frac{e^2}{3(4\p)^4 }
\int \limits_{0}^{\infty} {\rm d}s \, 
(3s^2 - s^3) (\J / 2) 
\, \coth (s\J/2) \, {\rm e}^{-s } \non \\
&=& \frac{e^2}{3(4\p)^4 }
~+~ \frac{e^2}{3(4\p)^4 }
\int \limits_{0}^{\infty} {\rm d}s \, s^3\,
\left\{ \frac{ (\J/2)^2 }
{ \sinh^2  (s\J /2 )} 
- {1 \over s^2} \right\}\, {\rm e}^{-s } ~.
\label{omega-III}
\eea

Combining the results 
(\ref{omega-I+II}) and (\ref{omega-III}), 
we finally obtain
\bea
 \O_{\rm two-loop}  (\J^2, 0)
&=&
\frac{e^2}{2(4\p)^4 } \non \\
&+&  \frac{e^2}{2(4\p)^4 }
\int \limits_{0}^{\infty} {\rm d}s \, s^3\,
( \J/ 2)^2
\left\{ 
\frac{ 1 }
{ \sinh^2  ( s\J / 2 )} 
- {1 \over (s\J/2) ^2 } \right\}
\, {\rm e}^{-s } ~.
\label{two-loop}
\eea
This should be compared with the one-loop 
result (\ref{one-loop-2}).
The second term in (\ref{two-loop})
is seen to be proportional to the third derivative
of the function (\ref{xi}), while 
$\O_{\rm one-loop}(\J^2, 0)$ 
was proportional to the first derivative 
of the same function $\xi$. It is natural to wonder:
Does this pattern persist at higher loops,
so that  the loop expansion for 
$\O(\J^2, 0)$ is equivalent to a derivative expansion 
of $\xi$?

The first term in (\ref{two-loop}) generates a 
non-vanishing $F^4$ quantum correction, while 
the other term produces $F^6$ and higher powers
of the field strength. The two-loop $F^4$ term 
was also computed in \cite{KM2} using the background
field formulation in $\cN=2$ harmonic superspace
\cite{BBKO} in conjunction with the results
of \cite{KM-heat}.
In spite of the  expectations of \cite{DS}, 
non-vanishing two-loop
$F^4$ quantum corrections also appear in some
$\cN=2$ superconformal theories 
in four dimensions \cite{KM3}.

\vskip.5cm

\noindent
{\bf Acknowledgements:}\\
We are grateful to the referee
for constructive comments and for pointing out misprints.
This work is supported in part by the Australian Research
Council and  UWA research grants.

\end{document}